\documentclass[12pt,preprint]{aastex} 

\renewcommand{\cite}{\citealp} 

\shorttitle{Stellar Populations in the UMa\,II dSph} 
\shortauthors{M. Dall'Ora et al.}

\begin{document}


\title{Stellar Archaeology in the Galactic halo with the Ultra-Faint Dwarfs: VI.
Ursa Major\,II\altaffilmark{1}}


\author{M. Dall'Ora\altaffilmark{2}, 
Karen Kinemuchi\altaffilmark{3}, 
Vincenzo Ripepi\altaffilmark{2},  
Christopher T. Rodgers\altaffilmark{4}, 
Gisella Clementini\altaffilmark{5}, 
Luca Di Fabrizio\altaffilmark{6}, 
Horace A.Smith\altaffilmark{7},  
Marcella Marconi\altaffilmark{2},  
Ilaria Musella\altaffilmark{2},  
Claudia Greco\altaffilmark{8},  
Charles A.Kuehn\altaffilmark{7},  
M\'arcio Catelan\altaffilmark{9},  
Barton J.Pritzl\altaffilmark{10}, 
Timothy C. Beers\altaffilmark{11}  
}

\altaffiltext{1}{Based on data collected at the 1.52 m telescope of the 
INAF-Osservatorio Astronomico di Bologna, Loiano,  Italy, at the 2.3 m telescope
of the Wyoming Infrared Observatory (WIRO) at Mt. Jelm, Wyoming, USA, and at the
1.8 m Perkins telescope of the Lowell Observatory, at Anderson Mesa, Flagstaff,
Arizona, USA.} 
\altaffiltext{2}{INAF, Osservatorio Astronomico di Capodimonte,
Napoli, Italy, dallora@na.astro.it, ripepi@na.astro.it, marcella@na.astro.it,
ilaria@na.astro.it} 
\altaffiltext{3}{NASA-Ames Research Center/Bay Area Environmental Research Institute
Mail Stop 244-30, P.O. Box 1, Moffett Field, CA 94035-0001; karen.kinemuchi@nasa.gov} 
\altaffiltext{4}{University of Wyoming, Department of Physics \& Astronomy,
Laramie, WY 82071, USA; crodgers@uwyo.edu} 
\altaffiltext{5}{INAF, Osservatorio Astronomico di Bologna, Bologna, Italy; gisella.clementini@oabo.inaf.it}
\altaffiltext{6}{INAF, Centro Galileo Galilei \& Telescopio Nazionale Galileo,
S. Cruz de La Palma, Spain; difabrizio@tng.iac.es} 
\altaffiltext{7}{Department of Physics and Astronomy, Michigan State University, East Lansing, MI 48824,
USA; smith@pa.msu.edu, kuehncha@msu.edu, beers@pa.msu.edu}
\altaffiltext{8}{Observatoire de Geneve, 51, ch. Des Maillettes, CH-1290
Sauverny, Switzerland; claudia.greco@obs.unige.ch} 
\altaffiltext{9}{Pontificia Universidad Cat$\rm{\acute{o}}$lica de Chile, Departamento de Astronom\'{\i}a y
Astrof\'{\i}sica,  Santiago, Chile; mcatelan@astro.puc.cl} 
\altaffiltext{10}{Department of Physics and Astronomy,  University of Wisconsin
Oshkosh, Oshkosh, WI 54901, USA; pritzlb@uwosh.edu} 
\altaffiltext{11}{Department of Physics \& Astronomy and JINA: Joint Institute for 
Nuclear Astrophysics, Michigan State University, East Lansing, MI 48824, USA}


\begin{abstract} We present a $B,V$ color-magnitude diagram (CMD) of the Milky Way dwarf satellite Ursa Major\,II (UMa\,II), spanning the 
magnitude range from $V \sim$ 15 to $V \sim $ 23.5 mag and extending over a 18 $\times$ 18 arcmin$^2$ area centered on the galaxy. Our 
photometry goes down to about 2 magnitudes below the galaxy's main sequence turn-off, that we detected at $V \sim 21.5$ mag.
We have discovered a \textit{bona-fide} RR Lyrae variable star in UMa\,II, which we use to estimate a conservative dereddened distance modulus for
the galaxy of $(m-M)_{0}=17.70 \pm 0.04 \pm 0.12$ mag, where the first error accounts for the uncertainties of the calibrated photometry, and the second reflects our lack of information on the metallicity
of the star. The
corresponding distance to UMa\,II is  $34.7 ^{+0.6} _{-0.7} (^{+2.0} _{-1.9})$ kpc.  Our photometry shows
evidence of a spread in the galaxy subgiant branch, compatible with a spread in metal abundance in the range between $Z$=0.0001 and $Z$=0.001.  Based on our estimate of
the distance, a comparison of the fiducial lines of the Galactic globular clusters
(GCs) M68 and M5 ([Fe/H]=$-2.27 \pm 0.04$ dex and  $-1.33 \pm 0.02$ dex, respectively),
with the position on the CMD of spectroscopically confirmed galaxy members, may suggest the existence of 
 stellar populations of different metal abundance/age in the central region of UMa\,II.
\end{abstract}





\section{Introduction} In the $\Lambda$-Cold Dark Matter ($\Lambda$-CDM) paradigm, larger 
galaxies are hierarchically assembled by merging of small cold dark matter fragments \citep{Diemand07,Springel05}, as also recently confirmed by the high-resolution simulations ``Aquarius" \citep{Lunnan11,Springel08}.
The idea is attractive when applied to our Galaxy, since it echoes the pioneering scenario suggested by \citet{SZ78} (SZ), 
in which the Milky Way (MW) was successively built up from small substructures. Indeed, first attempts to link the SZ scenario with the 
cosmological simulations foresaw the assembling of the Galactic halo starting from a number of satellites, and producing a number 
of tidal streams that should be (and are in fact) observed \citep{Bullock01}. The survivors of such a process should have the 
observational characteristics of the present day dwarf spheroidal galaxies (dSphs), which are old, metal-poor and gas deficient systems.

However, according to the $\Lambda$-CDM theory, the expected number of surviving fragments of 
the accretion process is one or two  orders of magnitude larger  than the observed number of MW ``bright" dSph companions. 
It was also pointed out that the  present-day ``bright" dwarfs may be the  \textit{wrong} remnants, as they 
could be instead tidal dwarf galaxies, with a non-primordial origin (e.g. \citealt{Metz08}). In any case this so-called 
``missing satellites problem" \citep{Moore99, Klypin99} has represented, so far, a major issue of the comparison
between  theoretical expectations and observational evidence \footnote{However, it is important to stress that 
cosmological simulations refer to cold dark matter minihaloes, which may or may not host a stellar population \citep{Bullock01}}. 
The discrepancy was
partially alleviated in the past few years by the discovery of  several
``ultra-faint" dwarf galaxies (UFDs) surrounding the MW, on the basis of the analysis of the
Sloan Digital Sky Survey (SDSS) data  (e.g., \citealt{Koposov09}, and references
therein).   The UFDs bring the number of known MW dSph satellites to $\approx
27$ (e.g. \citealt{Mateo98,Belokurov10} and references therein).  Since the sky
coverage of the SDSS is only $\sim 1/4$ of the celestial sphere, basic
statistical arguments suggest that tens of these  faint MW satellites are still
undiscovered, thus further narrowing the gap between theoretical expectations and
observational evidence unless, as suggested by \citet{Metz09}, the
distribution of these faint systems is not uniform and follows  instead a disc
around the MW.   
Whatever the case, in order to be ``proper" remnants of the \textit{primordial} Galactic halo contributors the
observed satellites should  host old stellar populations with properties
compatible with those of the MW halo. In particular, they should contain stars
as metal-poor  as $ [Fe/H]  < -3.0$ or $-4.0$ dex (``extremely metal-poor stars", \citealt{Beers05}) and RR Lyrae stars with
pulsation properties conforming to the Oosterhoff dichotomy \citep{Oosterhoff39}
observed  for field and cluster MW RR Lyrae variables (\citealt{Catelan09}, and
references therein).  Regarding both aspects the ``bright" MW dSph
satellites do not seem to be the possible primordial contributors to the
Galactic halo, as spectroscopic studies (e.g. \citealt{Helmi06}) show that
there are very few extremely metal-poor stars in most of the ``bright" MW dSph
satellites  (Tolstoy et al. 2009, and reference therein) and, on the other
hand, these galaxies are generally classified as ``Oosterhoff-intermediate" (e.g.
Carina: \citealt{Dallora03}; Fornax: \citealt{Bersier02})  because their
fundamental-mode RR Lyrae stars have average periods  ($\langle P_{ab} \rangle
$) intermediate between the Oosterhoff I (OoI; $\langle P_{ab} \rangle $ = 0.55
days)  and II (OoII; $\langle P_{ab} \rangle $ = 0.65 days) types observed for the MW field and cluster 
variables. The Ursa Minor dSph ($\langle P_{ab} \rangle $ = 0.638 days) 
and the Sagittarius ($\langle P_{ab} \rangle $ = 0.574 days) are the only
exceptions to this generally accepted ``rule'' among the ``bright"
dwarfs (\citealt{SCC09},  and references therein). 

The UFDs, instead, seem to possess both proper  
metal abundances, as shown by a number of spectroscopic studies (e.g.,
\citealt{Simon07,Kirby08,Frebel10}), as well as compatible properties of stellar populations and variable stars, as shown by our long-term project to monitor the UFD variable stars (see \citealt{Moretti09}, for a summary).  
In brief, the five galaxies we  have studied so far, namely, Bootes~I (\citealt{Dallora06}), Canes Venatici~I
(CVn\,I;  \citealt{Kuehn08}), Canes Venatici~II (CVn\,II; \citealt{Greco08}), Coma
Berenices (Coma; \citealt{Mus09}),  and Leo IV (\citealt{Moretti09}), contain RR
Lyrae stars with pulsation periods suggesting an Oo~II classification, with the
exception being CVn\,I, which appears to be Oosterhoff-intermediate. In this
respect, we remark that both total luminosity and global metallicity of CVn\,I make this object more likely a ``classical" dSph than a UFD (see e.g. Fig. 5 of \citealt{Kirby08}).
Continuing our study of variable stars in the Galactic UFDs  in this paper we present results for Ursa Major\,II (UMa\,II). 
 
UMa\,II (R.A.= 08$^h$51$^m$30$^s$, decl.= +63$^{\circ}$ 07$^{\prime}$ 48$^{\prime \prime}$,
J2000, \citealt{Zucker06}) was initially discovered as a candidate
star cluster in the Galactic halo by \citet{Grillmair06}, and only
afterwards recognized as a dwarf galaxy by \citet{Zucker06}.  
According to \citet{Simon07} and \citet{Munoz10} UMa\,II appears to
be a very elongated and extended object, likely undergoing tidal
disruption. This suggestion is supported by the existence of a
velocity gradient along the major axis of $8.4 \pm 1.4$ km $s^{-1}$
between eastern and western sides of the galaxy
\citep{Simon07}. This gradient shows the same direction of
elongation as discussed in \citet{Munoz10}. The galaxy was proposed
to be the progenitor of the so-called Orphan Stream
\citep{Fellhauer07}.  \citet{Zucker06} derived a half light radius of
$r_h \sim 13^{\prime}$ for the galaxy, and highlighted that the central part of
UMa\,II breaks up into three distinct clumps. Subsequently, 
\citet{Munoz10} on the basis of deep, wide-field Canada-France-Hawaii
Telescope photometry, showed that this apparent clumping is due to the
poor statistics of the bright stars, and disappears when deeper
data are used. Moreover, they showed that UMa\,II is more extended
than previously thought.  The same conclusion was reached by
\citet{Newberg10}, from the analysis of the Sloan Extension
for Galactic Understanding and Exploration (SEGUE,
\citealt{Yanny2009}) spectroscopic and SDSS/SEGUE photometric
data. These findings seem to rule out the link between UMa\,II and the
Orphan Stream. Spectroscopic studies disclosed the presence of very
metal-poor stars in the galaxy, with [Fe/H] $\sim -3.0$ dex, and
detailed abundance patterns consistent with those observed in the Galactic halo
(\citealt{Kirby08,Frebel10}). \citet{Martin08} derived new structural
parameters for UMa\,II, adopting a slightly different position for the
galaxy centre and a flatter shape with a slightly larger $r_h$,
compared to \citet{Zucker06}. In the following we use the structural
parameters derived by \citet{Zucker06}. However, our conclusions would
not be affected whether we used Martin et al.'s parameters, instead.

\section[]{Observations and Data Reduction}  

Time-series observations of UMa\,II
were obtained between January and March 2007 with three different telescopes. 
At the Wyoming Infrared Observatory (WIRO) 2.3m telescope, using  the WIRO
Prime Focus Camera, we obtained 40 images in Johnson $V$, 12 images in Johnson
$B$, and 22 images in Cousins $I$.  At the 1.8m Perkins telescope of the Lowell
Observatory,  we used the Perkins Re-Imaging SysteM (PRISM) instrument in
photometry mode to obtain 12 $V$, 4 $B$ and 4 $I$ images of UMa\,II.  Five
nights at the Loiano Observatory 1.5m telescope  provided 14 $B$ and 15 $V$
images using the Bologna Faint Object Spectrograph and Camera (BFOSC). The WIRO
observations cover the largest area with a field of view (FOV) of $17.8' \times
17.8'$ at the pixel scale of $0.55$ arcsec/pix, pointed on the center
coordinates of UMa\,II  derived by \citet{Zucker06}. The FOVs at the Lowell
and Loiano telescopes are slightly smaller, $13.65' \times 13.65'$ ($0.39$
arcsec/pix) and $13' \times 12.6'$ ($0.58$ arcsec/pix), respectively.  The total
area covered by our observations is $18.38' \times 18.26'$. This corresponds to
the central clump of UMa\,II, while the other two clumps reported by
\citet{Zucker06}  are both slightly outside the field covered by our
observations. The log of observations is provided in Table \ref{tab_log}. 

Datasets from each observatory were pre-reduced using standard
IRAF\footnote{IRAF is distributed by the National Optical Astronomical
Observatory, which is operated by the  Association of Universities for Research
in Astronomy, Inc., under cooperative agreement with the National Science
Foundation} techniques (bias subtraction and flat fielding).  Since we lack a
photometric calibration of the $I$-band data, observations in this band were not
considered in the subsequent analysis and only the $B,V$ data are presented in the paper.  Point-spread function (PSF) photometry
was performed with the DAOPHOT IV/ALLFRAME \citep{Stetson87, Stetson94} package.
After an accurate evaluation of the PSF of each individual frame,  a reference
image was built by averaging all the available frames and a source catalogue was
extracted from the stacked image. The source list was then passed to ALLFRAME,
which  performed homogeneous PSF photometry simultaneously on all images, thus
producing $b$ and $v$ instrumental magnitude catalogues for each telescope.
Typical internal photometric errors were of the order of $0.01-0.02$ mag at the horizontal branch (HB) level, 
on the averaged catalog, while the signal-to-noise ratio greatly changes between the individual exposures, 
due to the different telescope sizes and  very different exposure times adopted  at the various sites.

\begin{figure}  \includegraphics[width=8cm]{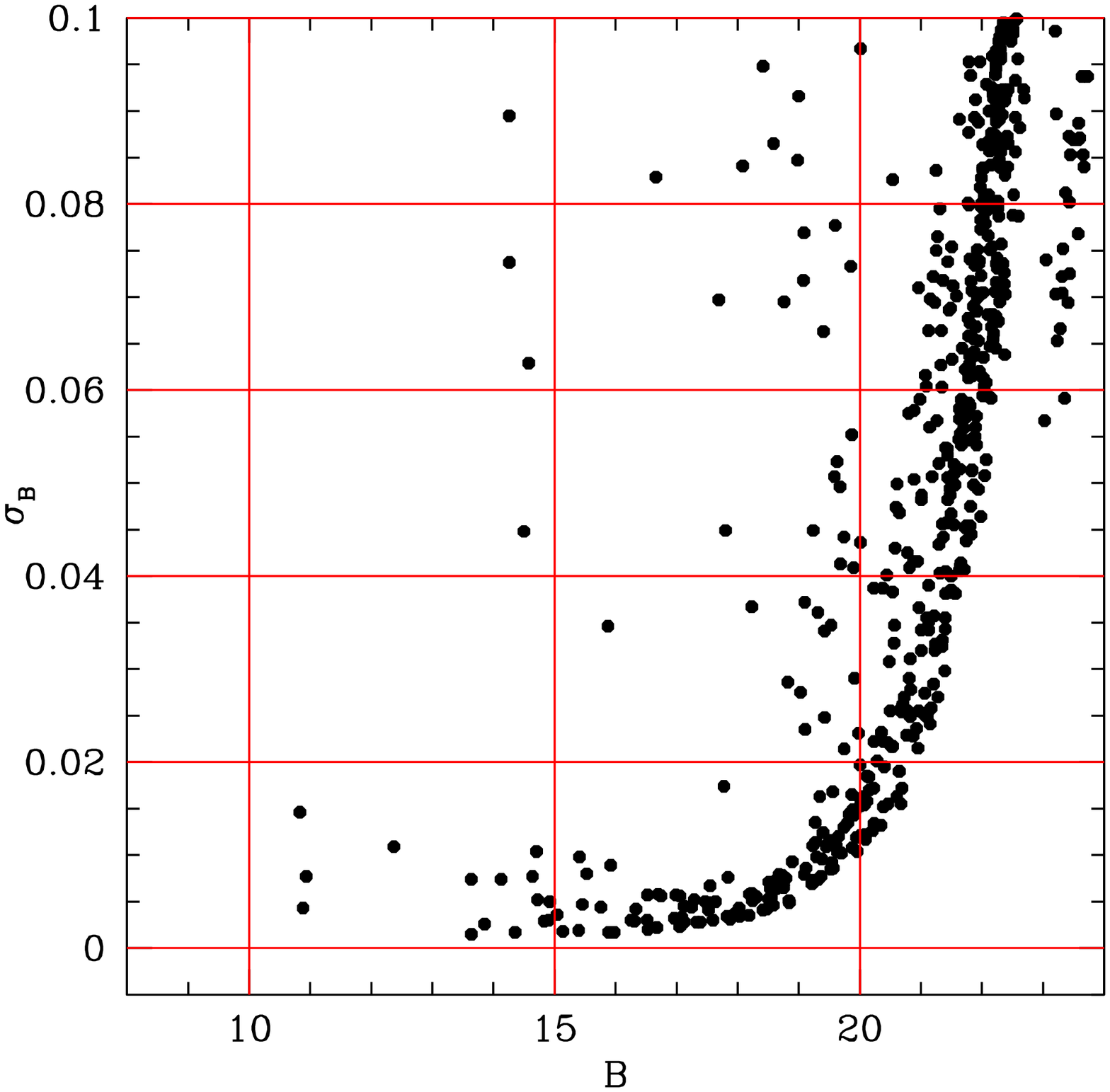}
\includegraphics[width=8cm]{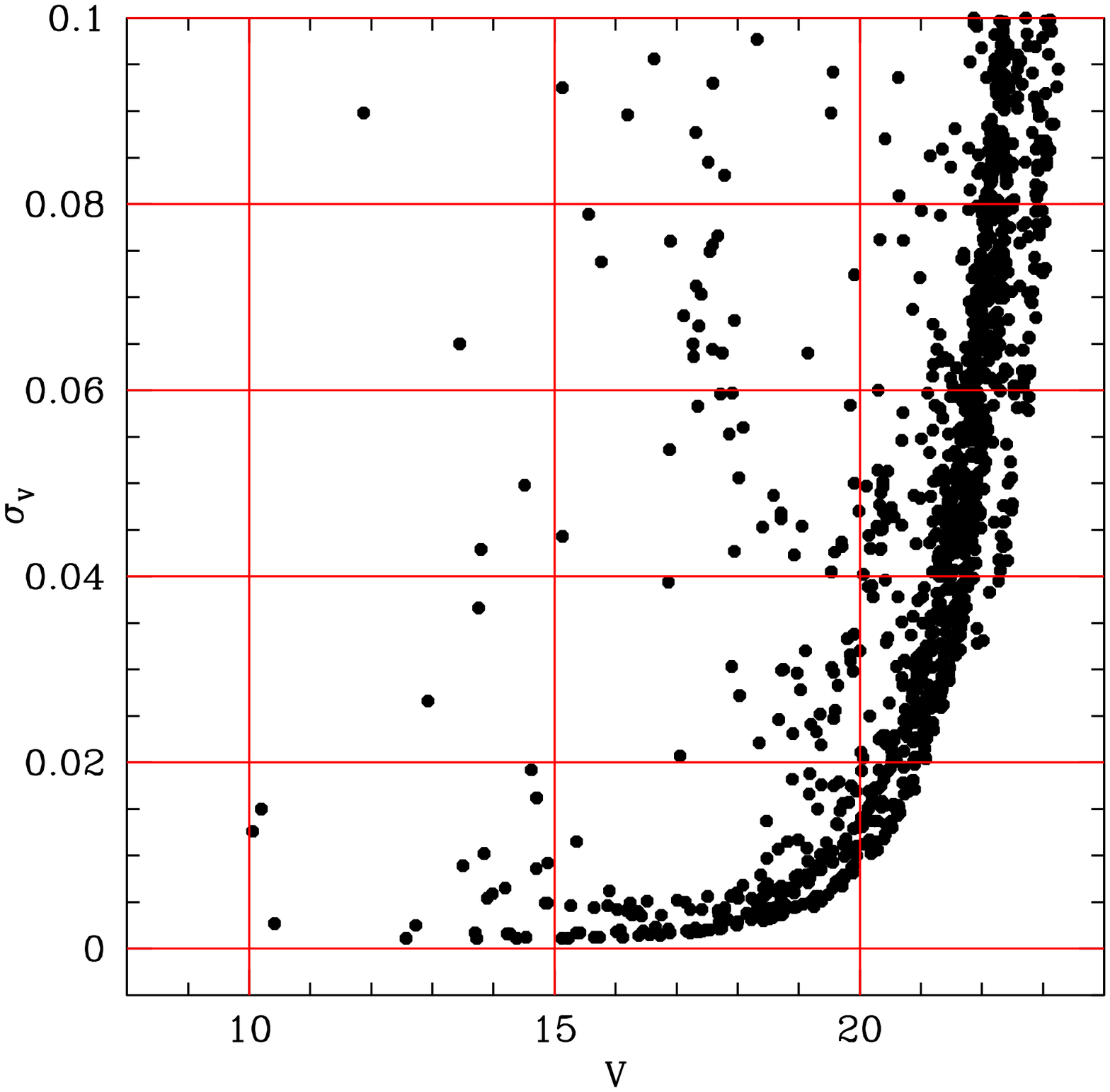} \caption{Photometric internal errors as a
function of the calibrated $B$ (left) and $V$ (right) magnitudes.} \label{errBV}
\end{figure}

To calibrate our photometry, additional $B,V$ observations of UMa\,II, along with
standard fields centered on the open clusters NGC 188 and NGC 7790, selected from the publicly available 
standard stars archive maintained by P.~B. Stetson \footnote{http://www4.cadc-ccda.hia-iha.nrc-cnrc.gc.ca/community/STETSON/standards/}, 
were obtained on a photometric night in 2008 January, at the 3.5m  Telescopio Nazionale Galileo in La Palma, Canary Islands. We used the standard La Palma extinction coefficients \footnote{http://www.ast.cam.ac.uk/$\sim$dwe/SRF/camc\_extinction.html} 
to derive the following calibration equations:

\begin{eqnarray}\nonumber 
B=b+(0.058\pm0.007)(B-V)+(26.400\pm0.020) \\  \nonumber  
V=v+(0.082\pm0.014)(B-V)+(26.141\pm0.010)   
\end{eqnarray}

where $B,V$ and $b,v$ are the standard and the instrumental magnitudes, respectively \footnote{The instrumental 
magnitude for a generic band $X$ is defined as $m_X=-2.5 \log(F_X)-K_X X$, where $F_X$ is the flux normalized to $1 sec$, 
$K_X$ is the extinction coefficient, and $X$ is the airmass.}.
The r.m.s. of these equations is of $\sim 0.02$ mag in both bands. 
These calibrations were used to define secondary standards in the field of UMa\,II. In particular, we selected 25 isolated local
standards with photometric errors less than $0.05$ mag which were visually
inspected to  remove any non-stellar objects.  The
remaining stars cover a range in color of $0.1 < B-V < 1.7$ mag.
Since the three telescopes define slightly  different photometric systems, the individual
datasets were calibrated separately, and for each filter a final master
catalogue was obtained by averaging the measurements of the individual telescopes. Typical 
uncertainties in calibrating our dataset on the TNG local standards were in the 
range $0.01-0.03$ mag. Figure \ref{errBV} shows the photometric internal error as a 
function of the calibrated magnitude, both in the $B$ and $V$ bands.

\section[]{Identification of Variable Stars and distance to the galaxy} 
\begin{figure}  \includegraphics[width=16.3cm]{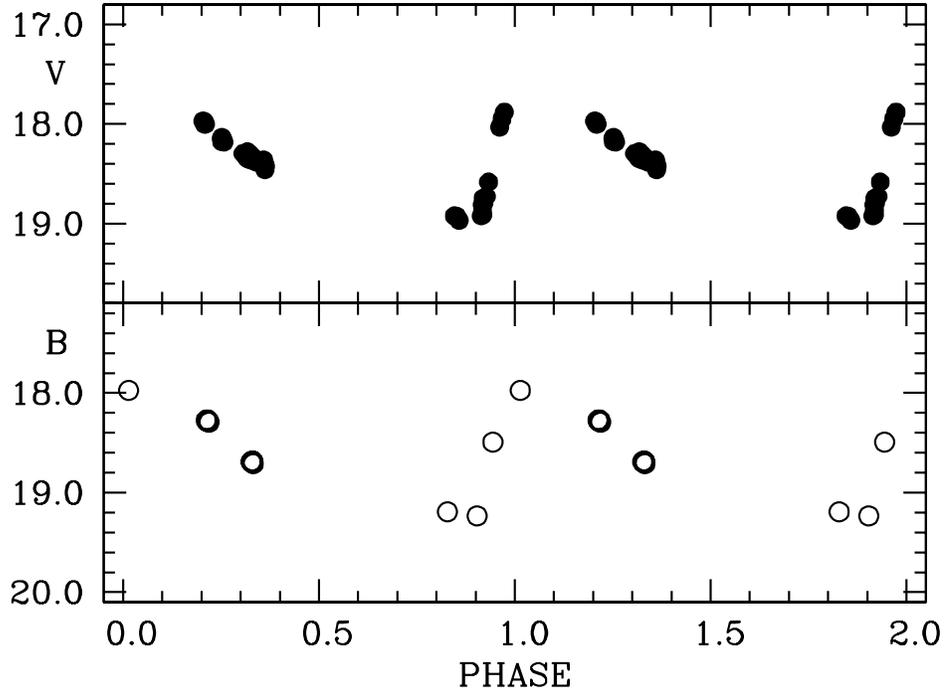} \caption{$B,V$ light
curves of the lonely bona-fide RR Lyrae variable, star V1, detected  in the
UMa\,II dSph. See text for details.} \label{lightcurve} \end{figure}

\begin{figure}  
\includegraphics[width=16.3cm]{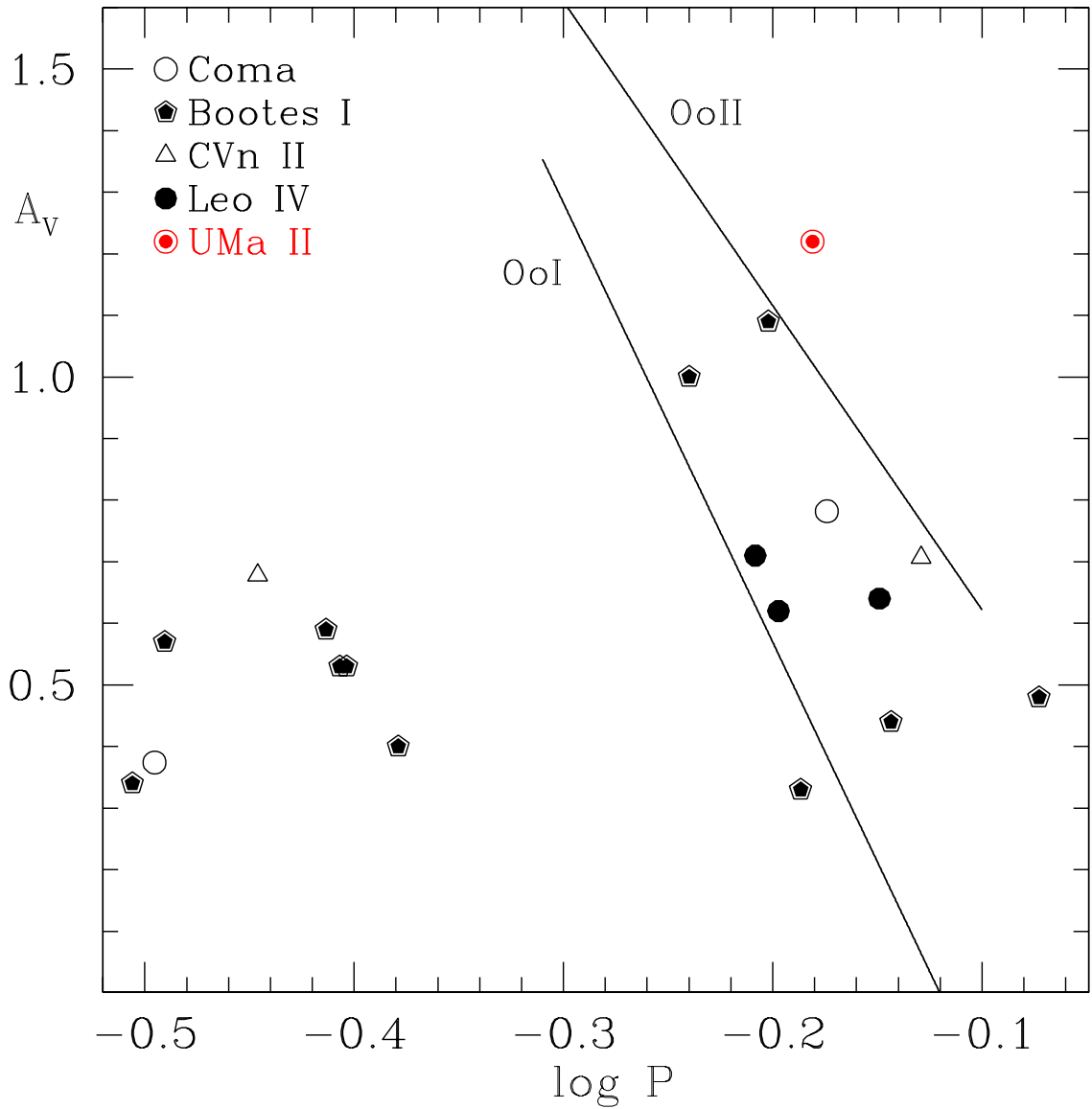} 
\caption{Period-amplitude diagram in the
$V$ band. The solid lines  show the positions of the Oo\,I and Oo\,II Galactic
GCs, according to \citet{CR00}.  
Different symbols correspond to RR Lyrae stars identified in  five
of the ultra-faint SDSS dwarfs studied so far for variability (namely Bootes\,I,
CVn\,II, Coma, Leo\,IV, and UMa\,II, from the present study.)
For UMa\,II we have plotted the amplitude of the truncated Fourier series best fitting the $V$ data (A$_{\rm V}$=1.22 mag). Results are unchanged if we use instead the amplitude corresponding to the difference of the observed minimum and maximum light data points (A$_{\rm V}$=1.09 mag).}
 \label{bailey}
\end{figure}

For each
star calibrated time-series photometry was obtained firstly by shifting the
instrumental magnitudes to the WIRO instrumental photometric system, and then calibrating the WIRO-aligned data to the Johnson photometric system. We omitted
the relative color term dependence since the relative color terms were small enough in
the observed color range of the RR Lyrae  stars ($ 0.2 < (B-V)_0 < 0.5$ mag) to be
considered negligible for our purposes. Search and identification of candidate
variable stars were performed with two different methods: 1)  the \citet{Stetson93} variability 
index, which compares the spread of individual measurements with the intrinsic
photometric error; 2) an \textit{ad hoc} procedure in which we first computed
the Fourier transforms (in the \citealt{Sch96}  formulation) of the stars having
at least 12 measurements in each photometric band, and then we averaged these
transforms to estimate the noise and calculated the signal-to-noise ratios.
Results from $B$ and $V$  photometries were cross-correlated, and all stars with
S/N$>$ 4 in both photometric bands were visually inspected, for a total of $\sim
200$ candidates. In particular, we checked all the stars around the galaxy's
horizontal branch and some of the stars in the blue
stragglers region of the CMD which might be pulsating variables of the SX Phoenicis
type. The light curves of the candidate variables were inspected by eye.  We
confirmed only one candidate variable, whose multiband light curves were analyzed with the
Graphical Analyzer of Time Series (GRaTiS, \citealt{Clementini00}) package,
obtaining a classification as RR Lyrae with pulsation period of 0.6593
days. We note that the time window of our data would imply an accuracy on the second decimal place of the derived period, 
however, we have used a period value to the forth decimal place to fold the light curves
of the RR Lyrae star, since this allowed us to significantly reduce the rms scatter of the truncated Fourier series best fitting the data. 
The star's time series data are provided in Table~2 and the pulsation characteristics 
(period, epoch of maximum light, amplitudes of the $V$, $B$ light variations, and 
intensity-averaged mean magnitudes) are summarized in Table~3. 
The internal accuracy of the average magnitudes of the RR Lyrae star is of $\pm$ 0.02 mag, as estimates from the 
 rms scatter of the models best fitting the light curves. 
 For the amplitudes we list both the values of the difference between minimum and maximum light data points, 
 and the values of  the truncated Fourier series best fit models of the light curves (in brackets). Corresponding uncertainties in the amplitudes are of the order of 0.1 mag.
Light curves are shown in
Figure~\ref{lightcurve}.  Figure \ref{bailey} shows the position of
the only
RR Lyrae in UMa\,II on the period-amplitude diagram defined by the RR Lyrae
stars of the Bootes\,I, CVn\,II, Coma, and Leo\,IV UFDs.  The pulsation period 
 and the position on the period-amplitude
diagram  both suggest the similarity of UMa\,II to Oosterhoff type II systems. This classification is unchanged 
if we round the period to two decimal places.
Figure \ref{cmd_ridge} shows the CMD of UMa\,II with the RR Lyrae star
marked by a cyan filled circle. The $B-V$ color of the RR Lyrae star 
($\langle B \rangle - \langle V \rangle$ = 0.22 mag)  appears to be slightly too blue for 
the star pulsation period. We will discuss this further at the end of the section.

We have estimated the distance to UMa\,II from its RR Lyrae, 
by adopting  the calibration of the
absolute $V$ magnitude of the RR Lyrae stars as a function of the
metallicity provided by \citet{Clementini03}  and \citet{Gratton04}, who derived $\Delta
M_V(RR) / \Delta[Fe/H] =0.214 \pm 0.047$ mag/dex from a sample of 100 RR Lyrae stars in the Large Magellanic Cloud (LMC) spanning the metallicity
range from [Fe/H]=$-$ 2.12 to $-$0.27 dex, and the RR Lyrae zero-point
$M_V =0.59$ mag at $[Fe/H]=-1.5$ dex, proposed by \citet{cc03}. We explicitly
note that, selecting a brighter zero-point of $M_V =0.54$ mag at
$[Fe/H]=-1.5$ dex, which is in agreement with the LMC 
distance modulus of $(m-M)_0 = 18.52$ mag \citep{Clementini03}, the kernel of our
analysis is not changed.  Adopting a metallicity of [Fe/H] = $-2.44 \pm 0.06$
dex for UMa\,II \citep{Kirby08}, an average apparent magnitude of  the
UMa\,II HB  of $\langle V_{\rm HB} \rangle = 18.39$ mag, as derived from the galaxy RR Lyrae star, and a
reddening of $E(B-V) = 0.096$ mag from the \citet{Schlegel98} maps, we
obtain a true distance modulus for the galaxy of $(m-M)_0 = 17.70 \pm
0.04$ mag, corresponding to $34.7^{+0.6}_{-0.7}$ kpc. The quoted
uncertainty takes into account only the uncertainties of the intrinsic
photometry, of the photometric calibration, and of the slope and zeropoint of the
RR Lyrae M$_V$ - [Fe/H] relationship. However, as pointed out by our refereee, the 
UMa\,II  red giants exhibit a considerable spread in metal abundance, ($\sigma_{\rm [Fe/H]}$=0.57 dex, according to 
\citealt{Kirby08}), that we should take into account since we lack a direct estimate of metallicity 
for the RR Lyrae star. This translates into an additional uncertainty of $\pm$0.12 mag on the distance modulus, which thus becomes
$(m-M)_0 = 17.70 \pm 0.04 \pm 0.12$ mag, corresponding to $34.7^{+0.6}_{-0.7}(^{+2.0} _{-1.9})$ kpc. We also note that, 
since the $M_V-[Fe/H]$ relationship is extrapolated for metallicities below  [Fe/H]=$-$ 2.1 dex and there are claims that  it
may not be unique and/or linear (e.g. \citealt{Caputo2000}), this error might still be an underestimate.
Our distance agrees within the uncertainties with the
value obtained by \citet{Zucker06} ($(m-M)_0 = 17.5 \pm 0.3$ mag)
based on isochrone-fitting of the main sequence and turn-off
regions. Our distance of 34.7 kpc also agrees with the predictions of
the heliocentric distance to UMa\,II by \citet{Fellhauer07}. These
authors simulated the orbit of UMa\,II obtaining for the galaxy a
distance of 34 kpc, which taken at the face value would support their
suggestion that UMa\,II is the progenitor of the Orphan
Stream. However, the  N-body simulations by \citet{Sales08} show that there is no obvious link
between the Orphan Stream and the orbit of UMa\,II.  

As mentioned previously the RR Lyrae star shows a rather blue $B-V$ color.
We do not have a clear explanation for this occurence, as 
the image of the star does not seem to be
contaminated by any companion, and the good agreement with the HB ridgeline of M68
as well as with the theoretical
models by \citet{Pietrinferni06} (see below) both in the $V,B-V$ and $B,B-V$ planes, seem to
support the notion that the star is not affected by blending with a hot
companion or by deceptive photometric errors. On the other hand, 
our $B$ light curve is based only on six different epochs, as the 12 exposures obtained at the WIRO telescope span a very short time interval and contribute in fact only 2 distinct epochs, thus the derived $\langle B \rangle$ magnitude may be somewhat uncertain. However, this uncertainty does not affect significantly our distance estimate. Our time-series photometry was referred to the WIRO data, with photometric transformations to the Johnson standard system given by

\begin{eqnarray}\nonumber 
B=b+0.115(b-v)+ 3.477 \\  \nonumber  
V=v-0.098(b-v)+ 3.223  
\end{eqnarray}

which translate into the $V$-$(B-V)$ transformation

\begin{eqnarray}\nonumber 
V=v-0.080(B-V)+ 3.342
\end{eqnarray}

Therefore, a shift as large as $0.4$ mag of the star $B-V$ color (thus pushing the variable on the ``red'' side of the Instability Strip), would affect the derived $\langle V \rangle$ luminosity by $\sim 0.03$ mag at most.

\section[]{The Stellar Population(s) of the UMa\,II UFD}

\begin{figure}  \includegraphics[width=16.3cm]{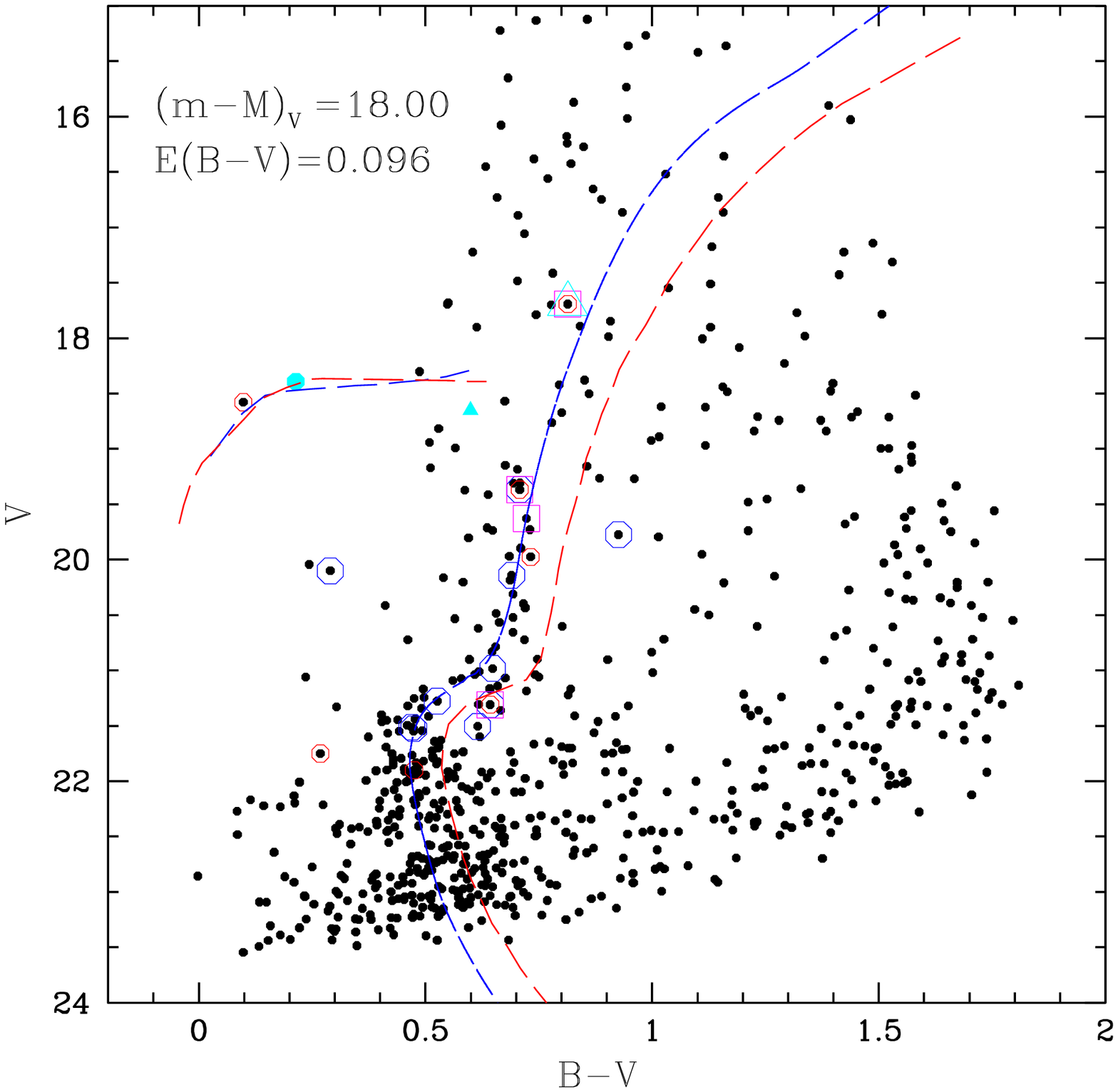} \caption{$V, B-V$
Color-Magnitude Diagram of UMa\,II with superimposed the ridgelines of the Galactic globular
clusters M68 (blue dashed line) and M5 (red dashed line).  Large blue, red,
magenta and cyan open symbols identify spectroscopically confirmed members of
UMa\,II  according to \citet{Martin07}, \citet{Simon07}, \citet{Kirby08}, and
\citet{Frebel10}, respectively. The cyan filled circle marks the galaxy's
bona-fide RR Lyrae star found in our study, while the cyan filled
triangle shows a suspected RR Lyrae in \citet{Simon07} study. This star  falls
outside the FOV of our observations, its magnitudes are taken from \citet{Simon07} and correspond to random phase values.} \label{cmd_ridge} \end{figure}

Figure \ref{cmd_ridge} shows our CMD of UMa\,II, where to minimize 
the contamination by non-point sources we only plot objects with DAOPHOT 
parameters $\chi \leq 1.4$ and $-0.3 \leq \textit{sharpness}
\leq +0.3$ and also considered only stars within the galaxy's half light radius ($r_h \sim 13$ arcmin, as worked out
from the preliminary values of distance, $d \sim 30$ kpc, and linear half light
radius $r_h \sim 120$pc, published by \citealt{Zucker06}). UMa\,II appears to be
quite elongated with an ellipticity of $\sim 0.5$, therefore  we adopted the
azimuthally averaged value of the half light radius (\citealt{Zucker06}). 
We explicitly note that  UMa\,II shows a rapidly varying ellipticity, 
with values ranging from $\epsilon = 0. 40$ in the center to $\epsilon
= 0.73$ near its half light radius \citep{Munoz10}, but this does not 
affect our conclusions about the stellar populations present in the galaxy.
The RR Lyrae star we have discovered in the galaxy is marked by a cyan filled circle, while the cyan filled triangle shows the position
of star SDSSJ084947.6+630830, which was suspected to be an RR Lyrae
variable by \citet{Simon07} on the basis of the observed radial velocity and spectral type variations. 
Unfortunately, this star falls outside our field of view. Nevertheless, we show  its position on our CMD, after transforming the SDSS $g,r$ magnitudes to
the Johnson  standard system with the relations by \citet{Fukugita96}, as it appears to be consistent 
with the RR Lyrae instability strip, thus supporting the \citet{Simon07} hypothesis. 
Blue, red, magenta and cyan open circles mark
spectroscopically confirmed members of UMa\,II according to \citet{Martin07},
\citet{Simon07}, \citet{Kirby08}, and \citet{Frebel10}, respectively. No
correction for contamination by Galactic field stars was made, as our attempt to
compare the PPM-Extended (PPMX) proper motions catalog \citep{Roeser10} of the spectroscopically
confirmed UMa\,II members with all the other stars present in the field did not show a clear separation.

UMa\,II is characterized by a very low average metal abundance $\langle$ [Fe/H]
$\rangle$ =$-2.44$ dex, an internal metallicity spread as large as $\sim 0.6$ dex (\citealt{Kirby08}), and $\alpha$ element abundances in agreement with those observed in the
Galactic halo ($[\alpha/Fe]=0.4$ dex (\citealt{Frebel10}, and references therein).
In Figure \ref{cmd_ridge} we have overplotted to the UMa\,II CMD the ridgelines of
the Galactic GCs M68 (blue dashed line, from \citealt{w94}) and M5 (blue solid
line, from \citealt{Sandquist96}), because these two GCs well encompass the
metallicity range found in the galaxy. Indeed, M68 is a metal-poor cluster, with
$[Fe/H]=-2.27 \pm 0.04$ dex \citep{carretta09}, and more recent determinations claim an
even lower metallicity of $[Fe/H]=-2.44 \pm 0.1$ dex
(\citealt{Simmons11}), hence
very similar to the average metallicity of UMa\,II; M5 is a metal-intermediate cluster,
with $[Fe/H]=-1.33 \pm 0.02$ dex \citep{carretta09}. 
The ridgelines were properly shifted to account for differences in distance and
reddening. For M68 we used the RR Lyrae mean apparent  \textit{V} magnitude 
$\langle V(RR) \rangle=15.60$ mag by  \citet{w94} and a reddening of $E(B-V)=0.05$ mag, as
available in the compilation by \citet{Harris96}. With these values we get an
apparent distance modulus of $15.11$ mag, by adopting the RR Lyrae
magnitude-metallicity calibration as above. For M5, we have adopted $(m-M)_0 =
14.44$ mag from \citet{Coppola11} and $E(B-V) = 0.03$ mag \citep{Harris96}.

\begin{figure}  \includegraphics[width=16.3cm]{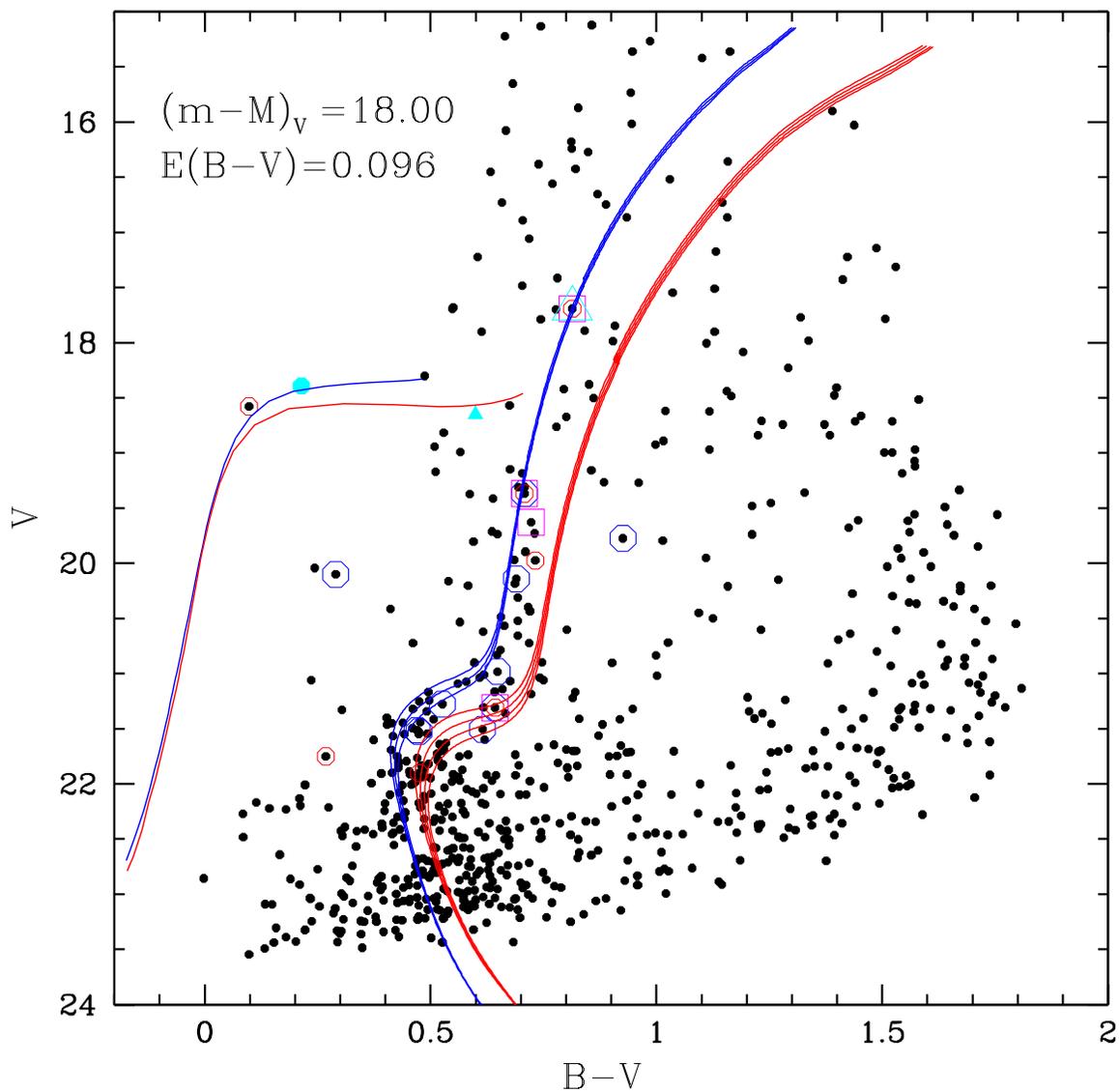} \caption{Isochrones from \citet{Pietrinferni06} overplotted to the galaxy CMD. We adopted $Z = 0.0001$ and $[\alpha/Fe]=0.4$ and ages of $12,
13$, and $14$ Gyrs  (blue solid lines), and $Z = 0.001$ and $[\alpha/Fe]=0.4$
and ages of $11, 12, 13$, and $14$ Gyrs  (red solid lines), respectively.}
\label{cmd_basti} 
\end{figure}

Our RR Lyrae-based distance to UMa\,II is supported by the good match of the M68
ridgelines and the UMa\,II subgiant branch (SGB), and by the very satisfactory
alignment, along both the M68 and M5 HBs, of three \textit{bona-fide} UMa\,II
stars, one of them being a spectroscopically confirmed member. 
The good match between few stars located at
$V$ in the range from $21.0$ to $21.5$ mag and $B-V$ color between $0.4-0.5$
and $0.7-0.8$ mag, two of which are spectroscopically confirmed members of
UMa\,II with the M5 ridgeline seems to suggest the presence of a second SGB. This is not surprising, since 
the quoted spectroscopic works clearly demonstrate the occurrence of stellar populations
with different metallicities in this UFD. Moreover, the galaxy \textit{bona-fide} main sequence appears
to have an observed width of $\approx 0.2$ mag in the range of luminosities
between $V \approx 22$ and $V \approx 23$ mag, while the width expected in that
range due to the photometric errors alone would be of the order of $\sigma_{B-V}
\approx 0.1$ mag (see Fig. \ref{errBV}). In the following, we will refer to the stars close to the M68
and M5 ridgelines as ``blue" and ``red" populations, respectively.
To check if the blue and red population are also compatible with a spread in the age, 
we have compared in Figure \ref{cmd_basti} the CMD of UMa\,II with the BaSTI (Bag of Stellar Tracks and Isochrones)
$\alpha$-enhanced isochrones by \citet{Pietrinferni06}. Following the
recent $\alpha$-enhanced abundances measured  by \citet{Frebel10}, we  have
adopted for the blue population the $\alpha-$enhanced models with $Z = 0.0001$
and $[\alpha/Fe]=0.4$ dex, corresponding to an iron-over-hydrogen ratio of
$[Fe/H]=-2.62$ dex and a global metallicity of $[M/H]=-2.27$ dex, and ages of $12$, $13$,
and $14$ Gyr. For the red population we have used the $\alpha-$enhanced models
with $Z = 0.001$ and $[\alpha/Fe]=0.4$ dex, corresponding to an iron-over-hydrogen
ratio of $[Fe/H]=-1.62$ dex and a global metallicity of $[M/H]=-1.27$ dex, with ages of
$11$, $12$, $13$ and $14$ Gyr, since M5 is thought to be $\sim 1$ Gyr younger than M68
(see \citealt{Marin09}). We stress that our selection of the isochrones is  entirely
based on the \textit{measured} metallicities. It is also important to note that
UMa\,II hosts a very metal-poor stellar component with [Fe/H] = $-3$ dex,
(\citealt{Frebel10}). Unfortunately, the BaSTI database does not contain such low
metallicities. 
We also explicitly note that the spectroscopically confirmed
member of UMa\,II at $V \approx 19.8$ mag and $(B-V) \approx 1$ mag has a measured
metallicity of $[Fe/H] = -1.8$ dex (SDSSJ085117.07+630347.3, \citealt{Martin07}) and appears to be definitely too
red for its metallicity, being redder than spectroscopically confirmed members
close to the M5 ridgeline and the $Z=0.001$ isochrones. The star could be
contaminated by a red companion, even if the values of its DAOPHOT structural
parameters $\chi$ and \textit{sharpness} do not support this explanation. Strangely, this star is not present in the 2MASS catalog, even if stars of similar luminosity are reported.

The isochrone fitting appears to be quite satisfactory (see Fig.~\ref{cmd_basti}) but, as commented by 
the referee, the small statistics does not allow us to prove the existence of two \textit{distinct} stellar
populations, with different age and/or metallicity in  UMa\,II.  It is worth to mention 
that \citet{Zucker06} suggested 
the presence of an age/metallicity spread in UMa\,II that could be accounted for by an 
an intermediate-metallicity stellar population with age of at least 10 Gyr.
We do not recover such a spread, but our data marginally suggest a gap between
the metal-poor and the metal-intermediate SGBs. If this gap
is not due to the poor statistics, UMa\,II may have experienced distinct bursts of
star formation during its star formation history, similar to those found in the Carina dwarf galaxy \citep{Monelli03}, for example. However, an alternative, intriguing explanation has been proposed by \citet{FrebelBromm10}, who
suggested that in the poor baryonic matter content of the UFDs, pristine (i.e.
Population III) core collapse (and maybe pair-instability) supernova events may
have suppressed further star formation and inhibited a homogeneous mixing of the
interstellar medium. According to this model, stars born close to the core
collapse SNe progenitors would be chemically enriched from the supernova yields,
while more distant stars would not. Therefore, the stellar
populations in UMa\,II could have the same age but different metallicity as arising
from differently enriched material. Alternatively, several Population III SNes could have enriched the environment but, 
given the shallow potential well of the UFDs, part of the enriched gas was lost \citep{Frebel11}. In either the case, 
this kind of galaxies should be regarded as  ``one-shot events". Our photometric data do not allow us to
distinguish between the two possible scenarios since the red population is
reasonably well described by isochrones in the range between 11 and 14 Gyrs.

\begin{figure} 
\includegraphics[width=16.2cm]{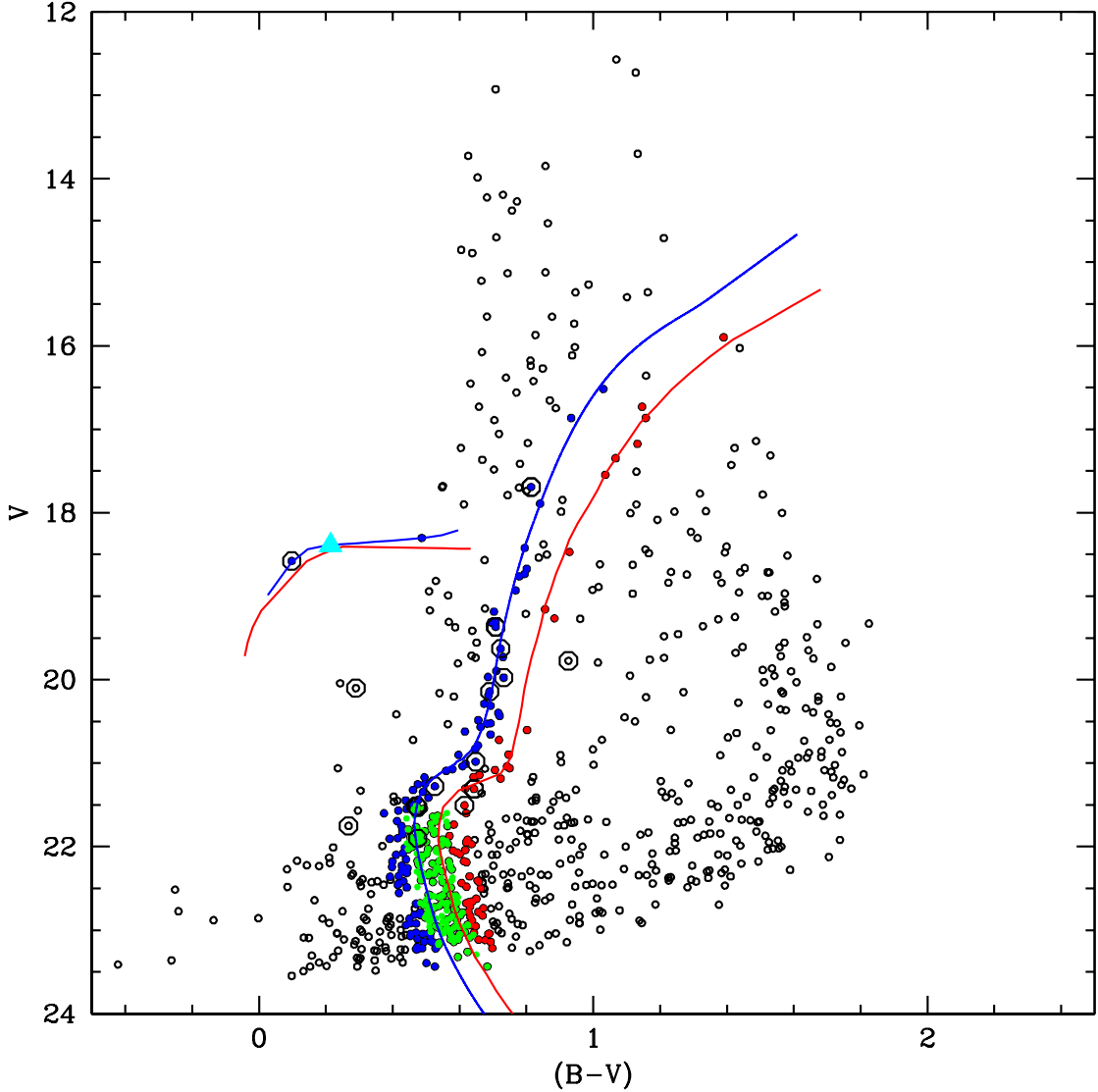}
\caption{$V, B-V$ Color-Magnitude Diagram of UMa\,II with the  ridgelines of the
Galactic GCs M68 (blue solid line) and M5 (red solid line) superimposed, which
we have used to select stars belonging to the two different populations that we
suggest to exist in UMa\,II. Blue and red dots mark the
stars within $\pm 0.05$ mag in $B-V$ for $V \leq$21.5 mag, and $\pm 0.10$ mag in
$B-V$ for $V > 21.5$ mag from the ridgelines of M68 and M5, respectively. Green dots show the sources with $V > 21.5$ mag which could belong to
either populations. Large open circles identify spectroscopically confirmed
members of UMa\,II according to \citet{Martin07}, \citet{Simon07}, and
\citet{Kirby08}.
The filled triangle marks the galaxy's bona-fide RR Lyrae star we detected in
our study.} \label{cmd_sel} \end{figure}

\begin{figure}[h] 

\includegraphics[width=9cm, height=9cm]{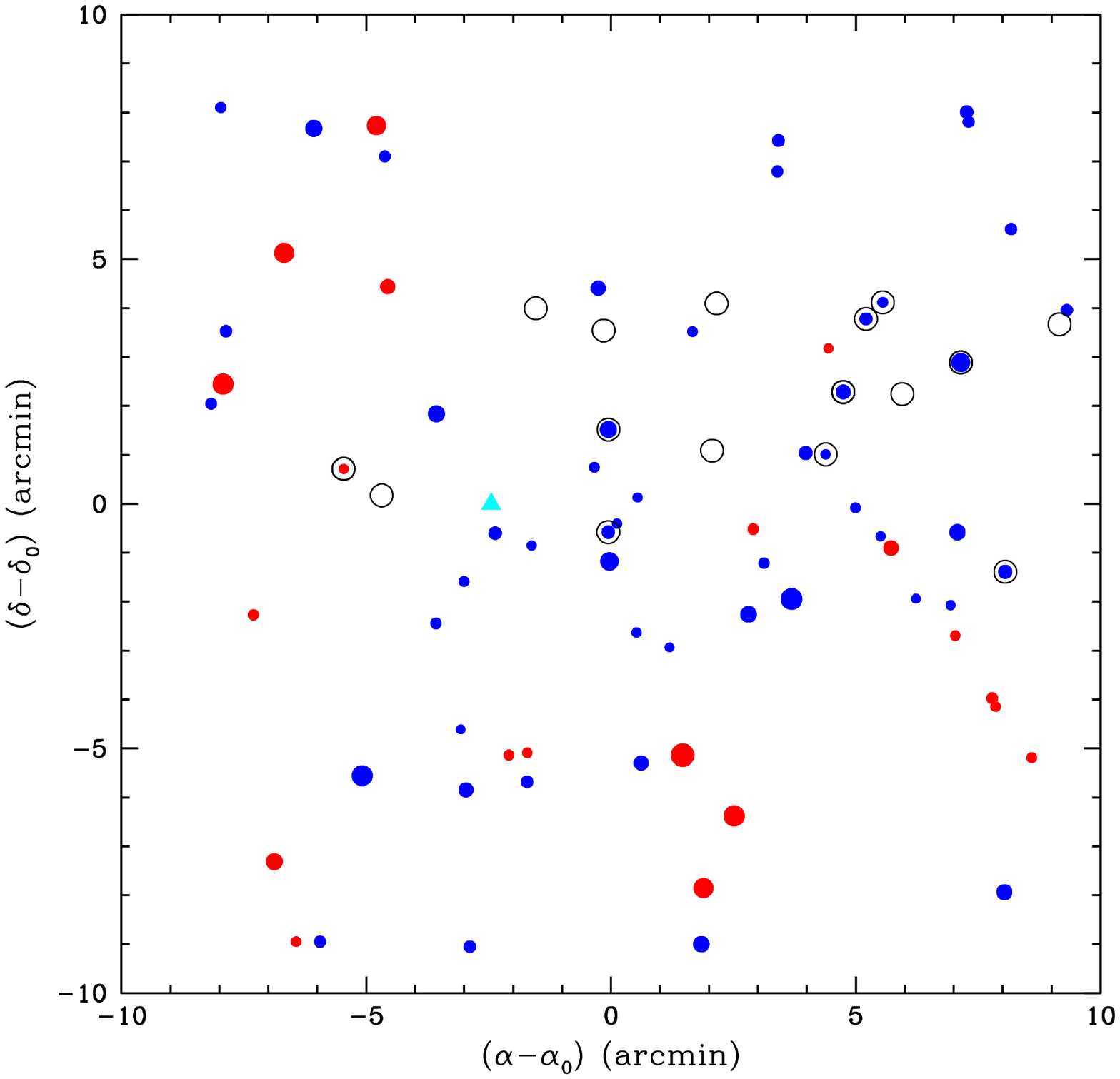} \includegraphics[width=9cm,
height=9cm]{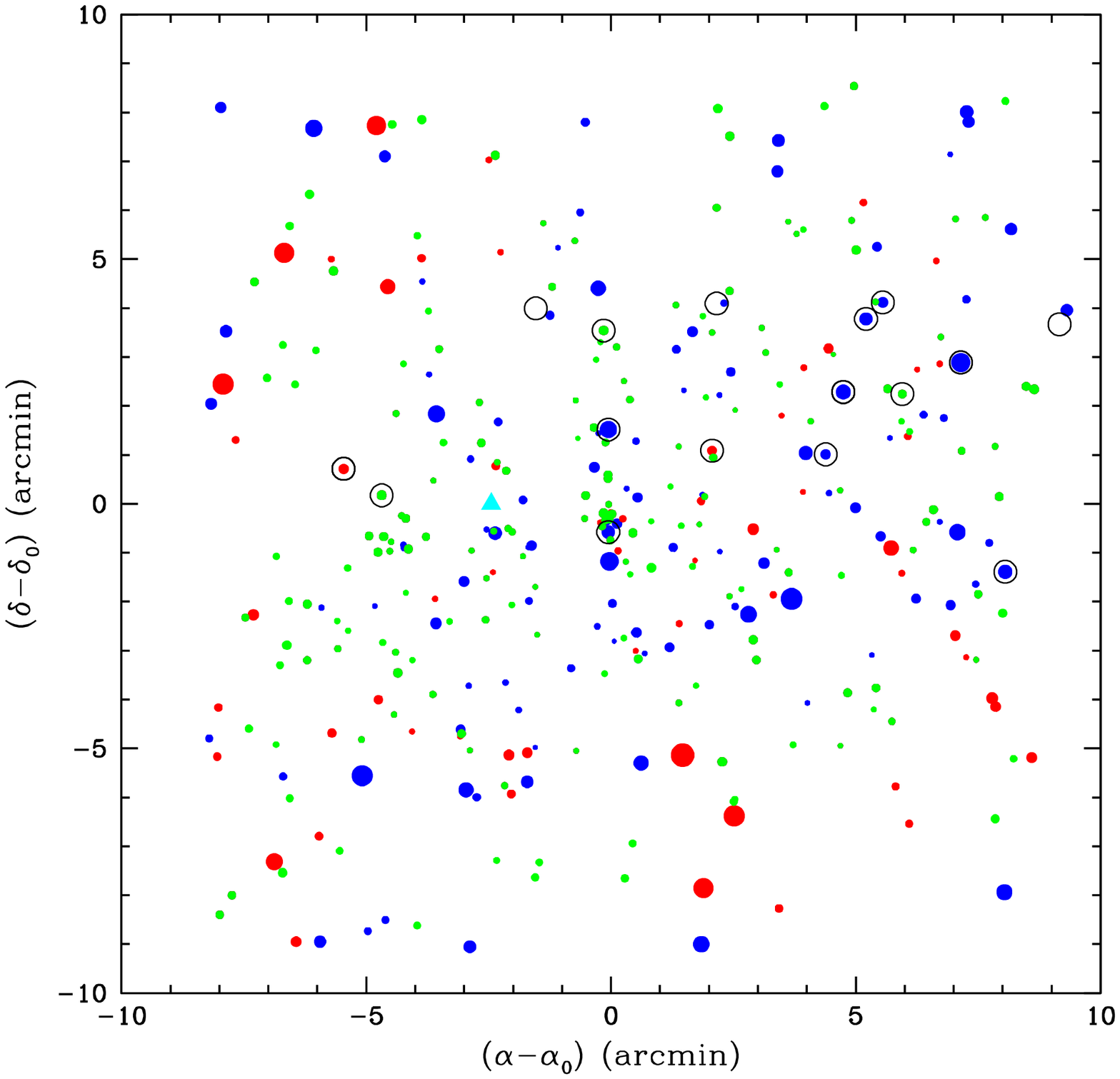}  
\centering \caption{ \textit{Upper panel:} Map of
sources with $V \leq$21.5 mag, within the UMa\,II half light radius, which we 
consider to belong to the galaxy according to the comparison with the ridgelines of M68 (blue filled circles) and M5 (red filled circles), or with
spectroscopically confirmed membership (black open circles). The symbol sizes are proportional to the star's brightness. 
{\it Lower panel:} Map of all sources within the UMa\,II half light radius,
which we consider to belong to the galaxy either according to the comparison with the
ridgelines of M68 (blue filled circles) and M5 (red filled circles), or with
spectroscopically confirmed membership (black open circles). Marked in green are
sources with $V >$21.5 mag which could belong to either populations.  
In both panels the RR Lyrae star is marked by a cyan filled triangle.} \label{map_sel}
\end{figure}

In order to verify whether the stellar populations show a different spatial
distribution in the field of view covered by our observations, we have used the
ridgelines of M68 and M5 to select stars possibly compatible with a blue and a red population, 
and plotted them  on the galaxy map with  different colors. This is shown in 
Figures \ref{cmd_sel} and \ref{map_sel}. In
particular, Figure \ref{cmd_sel}  illustrates our selection of the two components
by displaying as blue and red filled circles the sources, within the UMa\,II
half light radius,  which lie within $\pm0.05$ mag in $B-V$ for $V \le 21.5$ mag
and $\pm 0.10$ mag in $B-V$ for $V > 21.5$ mag from the ridgelines of M68 and
M5,  respectively. While the selection is rather straightforward for $V \le
21.5$ mag, at fainter magnitudes the two components cannot be totally 
separated. We have marked in green the sources with $V >21.5$ mag which could
belong to either population. Figure \ref{map_sel} shows the spatial  location of
these blue-, red-, and green-coded populations, in the FOV covered by our
observations, for sources with $V \le 21.5$ mag (upper panel)  and for all
sources (lower panel). The upper panel of Figure \ref{map_sel} shows that
sources brighter than $V = 21.5$ mag lying closer to the M68  ridgeline (blue
filled circles) are larger in number and seem to be more concentrated towards
the center of UMa\,II than sources closer to the M5  ridgeline (red filled
circles), but this could be an effect due to the poor statistics. On the other hand, the lower panel of Figure \ref{map_sel} shows that
sources fainter than $V =21.5$ mag, which could belong either to the M68-like or
to the M5-like population (green filled circles), do not show any obvious pattern.

\section[]{Conclusions} We have presented the first Johnson $V, B-V$ CMD of the
UMa\,II UFD galaxy and reported the detection of a \textit{bona-fide} RR Lyrae
star  in the galaxy, from which we have derived a true distance modulus of
$(m-M)_0=17.70 \pm 0.04 (\pm 0.12)$ mag (d=$34.7 ^{+0.6} _{-0.7} (^{+2.0} _{-1.9})$ kpc). The error quoted in the brackets takes into account our lack of knowledge of the metallicity of the RR Lyrae star. The pulsation period of this variable (P=0.6593 days) and its position 
in the period-amplitude (Bailey) diagram 
suggests the similarity of UMa\,II with Oosterhoff type II systems. A comparison of the observed CMD with the 
location of UMa\,II spectroscopically confirmed member stars
(taken from the catalogs of \citealt{Martin07,Simon07,Kirby08,Frebel10})  and
with the fiducial lines of the Galactic GCs M68 and M5 is compatible with a spread in age/metallicity of the stellar content
of the galaxy. The inner brightness profile of UMa\,II, 
as suggested by \citet{Munoz10} on the basis of deep CFHT photometry, could be due to the presence of multiple stellar generations, as also postulated in \citet{McConnachie07}.
However, the comparison with the isochrones gives only a coarse estimate of the age of the 
stellar populations, and no definite conclusion can be reached about 
a possible spread in age. On the other hand, when only the
comparison with the M68 and M5 ridgelines is considered, the hypothesis of an 
age difference of $\sim 1$ Gyr is favored. Since
the field covered by our photometry is smaller than the fields observed by the
\citet{Martin07},  \citet{Simon07}, \citet{Kirby08}, and \citet{Frebel10}
studies, our conclusions apply mainly to the central part of UMa\,II  monitored
by our data. A deeper, more accurate and more spatially extended photometry is
therefore highly desided to unveil all the secrets of the UMa\,II UFD.

\acknowledgments We thank an anonymous referee, for  helpful comments, which greatly improved both the scientific accuracy and the readability of this paper. We warmly thank Evan Kirby and Joshua Simon for providing us
identification and individual metallicities for member stars of the UMa\,II
UFD.  PRISM was developed with support from NSF (AST-0079541, K.A. Janes, P.I.)
with additional support from Boston University and Lowell Observatory. We also thank Benjamin Kelly and Laura Portscheller for having joined us in the first phase of this project. 
We acknowledge financial contribution from the Italian PRIN MUR 2007
``Multiple stellar populations in globular clusters: census, characterization
and origin"  P.I.: G. Piotto, from COFIS ASI-INAF I/016/07/0, and from the agreement ASI-INAF I/009/10/0. HAS thanks the NSF for support under grants AST-0607249 and AST-0707756.
MC acknowledges support by the Chilean Ministry for the
Economy, Development, and Tourism’s Programa Iniciativa Cient\'{i}fica
Milenio through grant P07-021-F, awarded to The Milky Way Millennium
Nucleus; by the BASAL Center for Astrophysics and Associated Technologies
(PFB-06); by the FONDAP Center for Astrophysics (15010003); by Proyecto
Fondecyt Regular \#1110326; and by Proyecto Anillo ACT-86.T.C.B. acknowledges partial funding of this work from grants PHY 02-16783 and PHY 08-22648: Physics Frontier Center/Joint Institute for Nuclear Astrophysics (JINA), awarded by the U.S. National Science Foundation.

\clearpage

\begin{table*} 
\caption{Observing Log. See text for details on the individual instruments.}\label{tab_log}
\vspace{0.5 cm} 
\begin{tabular}{cccccc}
\hline
\hline
{\rm Night}& {\rm Instrument} &	{\rm Filter}  & {\rm Exposure} & {\rm N. of images} & {\rm Median seeing}\\  
& & & (sec) & & (arcsec) \\ 
\hline
2007-01-16 & WIRO & V & 10 & 4 & 3.4\\ 
2007-01-17 & WIRO & V & 20 & 30 & 2.6\\ 
2007-01-17 & WIRO & V & 60 & 6 & 2.6\\ 
2007-01-17 & WIRO & B & 30 & 12 & 2.2\\ 
2007-02-22 & PRISM & V & 300 & 12 & 2.0\\ 
2007-02-22 & PRISM & B & 600 & 4 & 2.1\\ 
2007-03-12 & BFOSC & V & 1200 & 3 & 1.6\\ 
2007-03-12 & BFOSC & B & 1200 & 3 & 1.6\\
2007-03-13 & BFOSC & V & 1200 & 3 & 1.6\\ 
2007-03-13 & BFOSC & B & 1200 & 3 & 1.4\\ 
2007-03-14 & BFOSC & V & 1200 & 2 & 1.6\\ 
2007-03-14 & BFOSC & B & 1200 & 2 & 1.3\\ 
2007-03-15 & BFOSC & V & 1200 & 2 & 1.5\\ 
2007-03-15 & BFOSC & B & 1200 & 3 & 1.5\\ 
2007-03-16 & BFOSC & V & 1200 & 3 & 1.6\\ 
2007-03-16 & BFOSC & V & 1200 & 2 & 1.4\\ 
\hline   
\end{tabular} 
\end{table*}

\begin{table*}
\caption{$B,V$ Photometry of the RR Lyrae star detected in UMa\,II}
\vspace{0.5 cm} 
\begin{tabular}{ccccc}
\hline
\hline
{\rm HJD} & {\rm B}  & &{\rm HJD } & {\rm V}\\ 
{\rm ($-$2454117)} & {\rm (mag)}  & &{\rm ($-$2454116) } & {\rm (mag)}  \\
\hline
~~0.756437 &  18.27 & &  0.900689 &  18.92 \\
~~0.757167 &  18.27 & &  0.902199 &  18.81 \\
~~0.758047 &  18.28 & &  0.902819 &  18.87 \\
~~0.758747 &  18.27 & &  0.903249 &  18.91 \\
~~0.759427 &  18.29 & &  0.903679 &  18.75 \\
~~0.761087 &  18.29 & &  1.750797 &  17.97 \\
~~0.832037 &  18.69 & &  1.752317 &  17.98 \\
~~0.832677 &  18.69 & &  1.752877 &  17.98 \\
~~0.833317 &  18.70 & &  1.753467 &  18.00 \\
~~0.834037 &  18.68 & &  1.753997 &  18.00 \\
~~0.834687 &  18.69 & &  1.754927 &  18.00 \\
~~0.835567 &  18.71 & &  1.781567 &  18.15 \\
 36.764746 &  19.19 & &  1.782127 &  18.18 \\
 36.814344 &  19.23 & &  1.782657 &  18.13 \\
 36.841012 &  18.49 & &  1.783177 &  18.17 \\
 36.887420 &  17.97 & &  1.783707 &  18.17 \\
\hline			   
\end{tabular}

\label{t:table2}
\medskip
 Table~\ref{t:table2} is published in its entirety in the electronic edition of the Journal. 
 A portion is shown here
 for guidance regarding its form and content.
\end{table*}

\begin{table}
\tiny
\caption[]{Identification and properties of the RR Lyrae star detected in UMa\,II.}
\label{t:uma2}
     $$
         \begin{array}{lcclllcccccc}
	    \hline
            \hline
           \noalign{\smallskip}
           {\rm Name}&{\rm \alpha}&{\rm \delta}&{\rm Type}&~~~{\rm P}    &~~~{\rm Epoch}&{\rm \langle V\rangle}&~~{\rm N_V}&{\rm \langle B\rangle}&~~{\rm N_B}&{\rm A_V}  &{\rm A_B}\\
                     &{\rm (2000)}&{\rm (2000)}&          & ~{\rm (days)}& ($-$2450000) &           {\rm (mag)}&           &{\rm (mag)}           &           &{\rm (mag)}& {\rm (mag)}\\
                     &            &            &          &              &              &                      &           &                      &           &           &          \\
            \hline
            \noalign{\smallskip}
             {\rm V1}& 08:50:37.43   & +63:10:10.0   &{\rm RRab}& 0.6593       & 4153.8779    &     18.39            &   53      & 18.61                &    16     & \geq1.09(1.22)      & \geq 1.27(1.38)  \\
\hline															   
            \end{array}
	    $$
{\small Mean values are intensity-averaged mean magnitudes.}\\
\end{table}

\end{document}